\documentclass[twocolumn, switch]{article} % Method A for two-column formatting

\usepackage{preprint}

\usepackage{amsmath, amsthm, amssymb, amsfonts}

\usepackage[numbers,square]{natbib}
\usepackage[utf8]{inputenc}	% allow utf-8 input
\usepackage[T1]{fontenc}	% use 8-bit T1 fonts
\usepackage{xcolor}		% colors for hyperlinks
\usepackage[colorlinks = true,
            linkcolor = purple,
            urlcolor  = blue,
            citecolor = cyan,
            anchorcolor = black]{hyperref}	% Color links to references, figures, etc.
\usepackage{booktabs} 		% professional-quality tables
\usepackage{nicefrac}		% compact symbols for 1/2, etc.
\usepackage{microtype}		% microtypography
\usepackage{lineno}		% Line numbers
\usepackage{float}			% Allows for figures within multicol

\usepackage{lipsum}		%  Filler text

\usepackage{newfloat}
\DeclareFloatingEnvironment[name={Supplementary Figure}]{suppfigure}
\usepackage{sidecap}
\sidecaptionvpos{figure}{c}

\usepackage{titlesec}
\titlespacing\section{0pt}{12pt plus 3pt minus 3pt}{1pt plus 1pt minus 1pt}
\titlespacing\subsection{0pt}{10pt plus 3pt minus 3pt}{1pt plus 1pt minus 1pt}
\titlespacing\subsubsection{0pt}{8pt plus 3pt minus 3pt}{1pt plus 1pt minus 1pt}

\usepackage{tikz,xcolor,hyperref}

\definecolor{lime}{HTML}{A6CE39}
\DeclareRobustCommand{\orcidicon}{
	\begin{tikzpicture}
	\draw[lime, fill=lime] (0,0) 
	circle [radius=0.16] 
	node[white] {{\fontfamily{qag}\selectfont \tiny ID}};
	\draw[white, fill=white] (-0.0625,0.095) 
	circle [radius=0.007];
	\end{tikzpicture}
	\hspace{-2mm}
}
\foreach \x in {A, ..., Z}{\expandafter\xdef\csname orcid\x\endcsname{\noexpand\href{https://orcid.org/\csname orcidauthor\x\endcsname}
			{\noexpand\orcidicon}}
}
\title{RegWSI: Whole Slide Image Registration using Combined Deep Feature- and Intensity-Based Methods: Winner of the ACROBAT 2023 Challenge}

\def\reviewcolor{black}

\usepackage{authblk}

\author[1,2\thanks{\tt{wodzinski@agh.edu.pl}}]{Marek Wodzinski\orcidA{}}
\author[1]{Niccol\`{o} Marini\orcidB{}}
\author[1,3]{Manfredo Atzori\orcidC{}}
\author[1,4]{Henning M\"{u}ller\orcidD{}}

\affil[1]{Institute of Informatics, University of Applied Sciences Western Switzerland, Sierre, Switzerland}
\affil[2]{Department of Measurement and Electronics, AGH University of Kraków, Krakow, Poland}
\affil[3]{Department of Neuroscience, University of Padova, Padova, Italy}
\affil[4]{Medical Faculty, University of Geneva, Geneva, Switzerland}

\affil[-]{DOI: https://doi.org/10.1016/j.cmpb.2024.108187}

\begin{document}

\twocolumn[ % Method A for two-column formatting
  \begin{@twocolumnfalse} % Method A for two-column formatting
  
\maketitle

\begin{abstract}

\subsection*{Background and objective}
The automatic registration of differently stained whole slide images (WSIs) is crucial for improving diagnosis and prognosis by fusing complementary information emerging from different visible structures. It is also useful to quickly transfer annotations between consecutive or restained slides, thus significantly reducing the annotation time and associated costs. Nevertheless, the slide preparation is different for each stain and the tissue undergoes complex and large deformations. Therefore, a robust, efficient, and accurate registration method is highly desired by the scientific community and hospitals specializing in digital pathology.

\subsection*{Methods}
We propose a two-step hybrid method consisting of (i) deep learning- and feature-based initial alignment algorithm, and (ii) intensity-based nonrigid registration using the instance optimization. The proposed method does not require any fine-tuning to a particular dataset and can be used directly for any desired tissue type and stain. The registration time is low, allowing one to perform efficient registration even for large datasets. The method was proposed for the ACROBAT 2023 challenge organized during the MICCAI 2023 conference and scored 1st place. The method is released as open-source software.

\subsection*{Results}

The proposed method is evaluated using three open datasets: (i) Automatic Nonrigid Histological Image Registration Dataset (ANHIR), (ii) Automatic Registration of Breast Cancer Tissue Dataset (ACROBAT), and (iii) Hybrid Restained and Consecutive Histological Serial Sections Dataset (HyReCo). The target registration error (TRE) is used as the evaluation metric. We compare the proposed algorithm to other state-of-the-art solutions, showing considerable improvement. Additionally, we perform several ablation studies concerning the resolution used for registration and the initial alignment robustness and stability. The method achieves the most accurate results for the ACROBAT dataset, the cell-level registration accuracy for the restained slides from the HyReCo dataset, and is among the best methods evaluated on the ANHIR dataset.

\subsection*{Conclusions}

The article presents an automatic and robust registration method that outperforms other state-of-the-art solutions. The method does not require any fine-tuning to a particular dataset and can be used out-of-the-box for numerous types of microscopic images. The method is incorporated into the DeeperHistReg framework, allowing others to directly use it to register, transform, and save the WSIs at any desired pyramid level (resolution up to 220k x 220k). We provide free access to the software. The results are fully and easily reproducible. The proposed method is a significant contribution to improving the WSI registration quality, thus advancing the field of digital pathology.

\end{abstract}
\vspace{0.3cm}
\vspace{0.3cm}
\end{@twocolumnfalse} % Method A for two-column formatting
] % Method A for two-column formatting

%%%%%%%%%%%%%%%  Main text   %%%%%%%%%%%%%%%
% \linenumbers

\section{Introduction}
\label{sec:introduction}

The automatic registration of whole slide images (WSIs) is crucial for several downstream applications in digital pathology: (i) the multimodal diagnosis and prediction using artificial intelligence (AI) techniques~\cite{ANHIR1}, (ii) the annotation transfer between consecutive or restained slides to decrease the annotation time, and thus the related costs~\cite{weitz2023increasing}, (iii) the 3-D reconstruction of consecutive slides to create the volumetric representation~\cite{song20133d}.

It is challenging to propose an automatic, robust, and efficient WSI registration method that does not require fine-tuning or hyperparameter adaptation to a particular dataset. There are several associated challenges: (i) the slides are acquired in various medical centers worldwide using different equipment, stains, and acquisition protocols, resulting in high heterogeneity and varying quality, (ii) no assumption about the relative orientation of the registered WSIs can be made leading to the necessity of large receptive fields of the learning-based methods, (iii) the requirement of relatively low computational complexity which is in contrast to the resolution of the WSI images (that is usually at the level of more than 200k x 200k pixels).

\begin{figure*}[!htb]
    \centering
    \includegraphics[width = 0.85\textwidth]{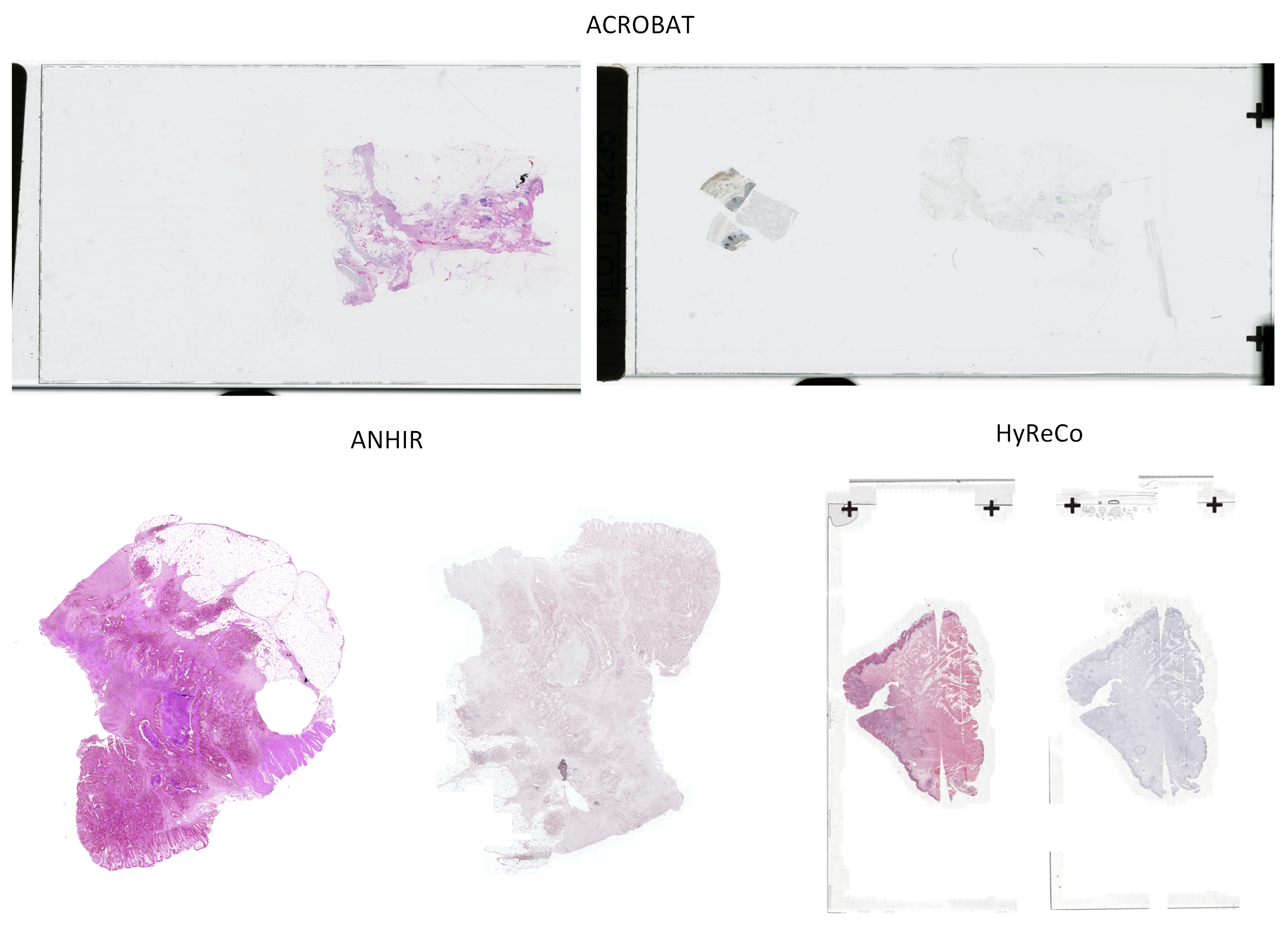}
    \caption{Exemplary registration pairs from the ACROBAT, ANHIR, and HyReCo datasets. Note the clinical quality of the ACROBAT samples (without any initial preprocessing and numerous artifacts), large initial misalignment in the ANHIR dataset, and the good quality of the HyReCo samples.}
    \label{fig:dataset}
\end{figure*}

The heterogeneity of data in digital pathology is enormous. The images are acquired in biopsy or resection procedures from nearly any part of not only the human body, but also other species. The samples are being prepared according to standards from different acquisition centers and digitized using various available scanners. The quality of the samples varies strongly with a large number of potential artifacts~\cite{taqi2018review,janowczyk2019histoqc}. All these factors lead to the requirement of high generalizability of the digital pathology algorithms. \textcolor{\reviewcolor}{A visualization} of registration pairs acquired in different medical centers is \textcolor{\reviewcolor}{shown in Figure~\ref{fig:dataset}}. Unfortunately, the current state-of-the-art methods in WSI registration still struggle with the necessity of fine-tuning for learning-based solutions or the requirement of hyperparameters adjustment for the classical, iterative methods.

In digital pathology, in contrast to radiology, no assumption about the initial orientation of the registered images can be made. The same observation applies to the \textcolor{\reviewcolor}{initial spatial offset}. Therefore, traditional intensity-based methods usually fail making the initial alignment of WSIs considerably challenging. The problem is even more impactful for the learning-based methods resulting in the requirement of large receptive fields, limiting the usefulness of CNN-based architectures.

The resolution of WSIs makes it difficult to perform efficient registration. The WSIs are usually saved in pyramidal format with the highest resolution above 200k x 200k pixels. That makes the WSIs orders of magnitude larger than the high-resolution computed tomography (CT) or magnetic resonance (MR) 3-D volumes, leading to significant computational complexity. Considering that the registration is usually the initial processing step performed before the downstream task, the registration time should be relatively low. For the consecutive slides the problem can be mitigated by reasonable downsampling, however, for the restained slides, the high resolution should be maintained to enable the cell-level registration accuracy.

\textbf{Contribution: } In this article, we propose an automatic, robust, efficient, and accurate method for the registration of whole slide images (WSIs). We evaluate the method using three open datasets and show a significant improvement with respect to the current state-of-the-art in WSIs registration. We openly release the method source code and incorporate it in the DeeperHistReg framework allowing other researchers to directly use it in their research~\cite{source_code,deeperhistreg}. The method has superior generalizability, it does not require any fine-tuning or retraining to a particular dataset. Moreover, it can be used to register other kinds of microscopic images, without any changes to the parameters setup.

\section{Related Work}
\label{sec:related_work}

There are numerous contributions in the field of automatic WSI registration, both in terms of open data availability, as well as methodological advances. 

\textcolor{\reviewcolor}{Currently there are} three major open datasets dedicated to evaluating advances in WSI registration. The first and probably the most recognized is the ANHIR dataset introduced during a challenge organized jointly with the IEEE ISBI 2019 conference~\cite{ANHIR1}. The dataset is heterogeneous, containing several tissue types stained with various dyes. However, the dataset is of relatively good quality, without significant artifacts or distortions, which is rare in practical scenarios. The second dataset was released during the ACROBAT challenge organized jointly with the MICCAI conference~\cite{ACROBAT1,ACROBAT2}. The dataset consists of several hundred breast cancer clinical slides acquired using various stains. The quality of the slides varies strongly, from good quality ones to cases with foreign bodies, low contrast, and missing or additional tissues. The third notable dataset is called HyReCo~\cite{HyReCo1,HyReCo2} and contains both consecutive and restained slides. The dataset consists of high-resolution and good-quality sections and enables researchers to evaluate the influence of the registration resolution on the restained samples. Other big datasets could be useful for evaluating WSI registration, like MSHIR~\cite{lin2023end} or COMET~\cite{awan2023deep}, however, up-to-date they remain closed to the scientific community.

\textcolor{\reviewcolor}{Medical image registration is a research area with a significant number of notable contributions~\cite{sotiras2013deformable,hering2022learn2reg,mok2020large,balakrishnan2019voxelmorph,klein2009elastix,haskins2020deep,mahapatra2018deformable,marstal2016simpleelastix,glocker2011deformable,heinrich2012mind,maes2003medical}}. However, due to the challenges specific to the WSIs, the general state-of-the-art registration methods usually fail~\cite{ANHIR2}. Therefore, contributions proposed specifically to WSIs are required. The two most successful methods are named HistokatFusion~\cite{lotz2019robust} and VALIS~\cite{VALIS}. Both of them are based on classical, iterative image registration. The HistokatFusion is ready-to-use closed software implementing both initial and nonrigid intensity-based iterative procedures with normalized gradient fields (NGF) as a similarity measure and B-Splines as the deformation model. The method was the winner of the ANHIR challenge and was among the best-performing ones during the ACROBAT challenge~\cite{ACROBAT1,ACROBAT2}. The VALIS library is fully operational software using SimpleITK as the registration backend. The main limitation of VALIS is a significant computational complexity limiting the usefulness to consecutive slides, with limited applicability to registering the restained ones. There are also numerous other scientific contributions, however, not directly useful in practice. Among the classical ones, it is worth mentioning: (i) the UPENN method based on the "Greedy" tool dedicated to diffeomorphic registration~\cite{venet2019accurate,venet2021accurate}, (ii) the multistep AGH method based on the Demons algorithm with modality independent neighborhood descriptor~\cite{wodzinski2021multistep}, (iii) the intensity-based registration driven by unsupervised classification of structural similarity~\cite{song2013unsupervised}, (iv) the patch-based nonrigid registration method enabling registration of WSIs at any resolution~\cite{lotz2015patch}, (v) a combined feature- and quad-tree based method for robust affine alignment~\cite{marzahl2021robust}.

There are several deep learning-based contributions to the WSI registration. The TUB method based on a modified volume tweening network was among the first ones~\cite{zhao2019unsupervised} and scored considerably well in the ANHIR challenge. The method was not on the podium, however, confirmed that the learning-based solutions can significantly speed up the registration process. Another worth mentioning contribution is the DeepHistReg method that introduced a patch-based approach to the learning-based WSI registration, allowing one to perform the learning-based registration at any desired resolution~\cite{wodzinski2021deephistreg,wodzinski2020unsupervised}. One of the most accurate solutions is based on CNN guided by structural features, with a fully unsupervised training pipeline~\cite{ge2022unsupervised}. Another work attempted to use data-driven descriptors learned by CNN architecture and compared them to the hand-crafted descriptors~\cite{awan2023deep}. Additionally, the proposed method was combined with the HistokatFusion deformable registration resulting in significant improvement. Recently, several researchers attempted to address the initial alignment of WSIs. Several works used transformers to increase the receptive field~\cite{pyatov2022affine,pyatov2022tahir}, thus solving one of the limitations of CNN-based affine registration networks~\cite{wodzinski2020learning}. Another work proposed an end-to-end affine registration method by utilizing weak annotations~\cite{lin2023end}. The researchers reformulated the problem into the classification of the initial misalignment to reduce the requirement of a large receptive field of the following affine registration network.

There are several limitations of the current state-of-the-art registration algorithms: (i) the majority of the methods are evaluated using just a single dataset, rarely coming from different acquisition centers, (ii) the algorithms require fine-tuning or hyperparameters adjusting whenever samples from new datasets are going to be registered, (iii) the initial alignment is not robust resulting in numerous failures, and (iv) the source code, which is also rarely released, usually do not allow other researchers to directly replicate the experiments or use the methods in practical applications.

In this work, we solve the limitations: (i) we evaluate the proposed method on several datasets acquired in different medical centers and show good stability and generalizability, (ii) the proposed method does not require any fine-tuning, retraining or parameter adjustment to new datasets, (iii) the initial alignment method works correctly for almost any case, and (iv) we openly release the source code and incorporate it in the DeeperHistReg framework allowing others to use the algorithm in research and practical applications.

\begin{figure*}[!htb]
    \centering
    \includegraphics[width = 0.95\textwidth]{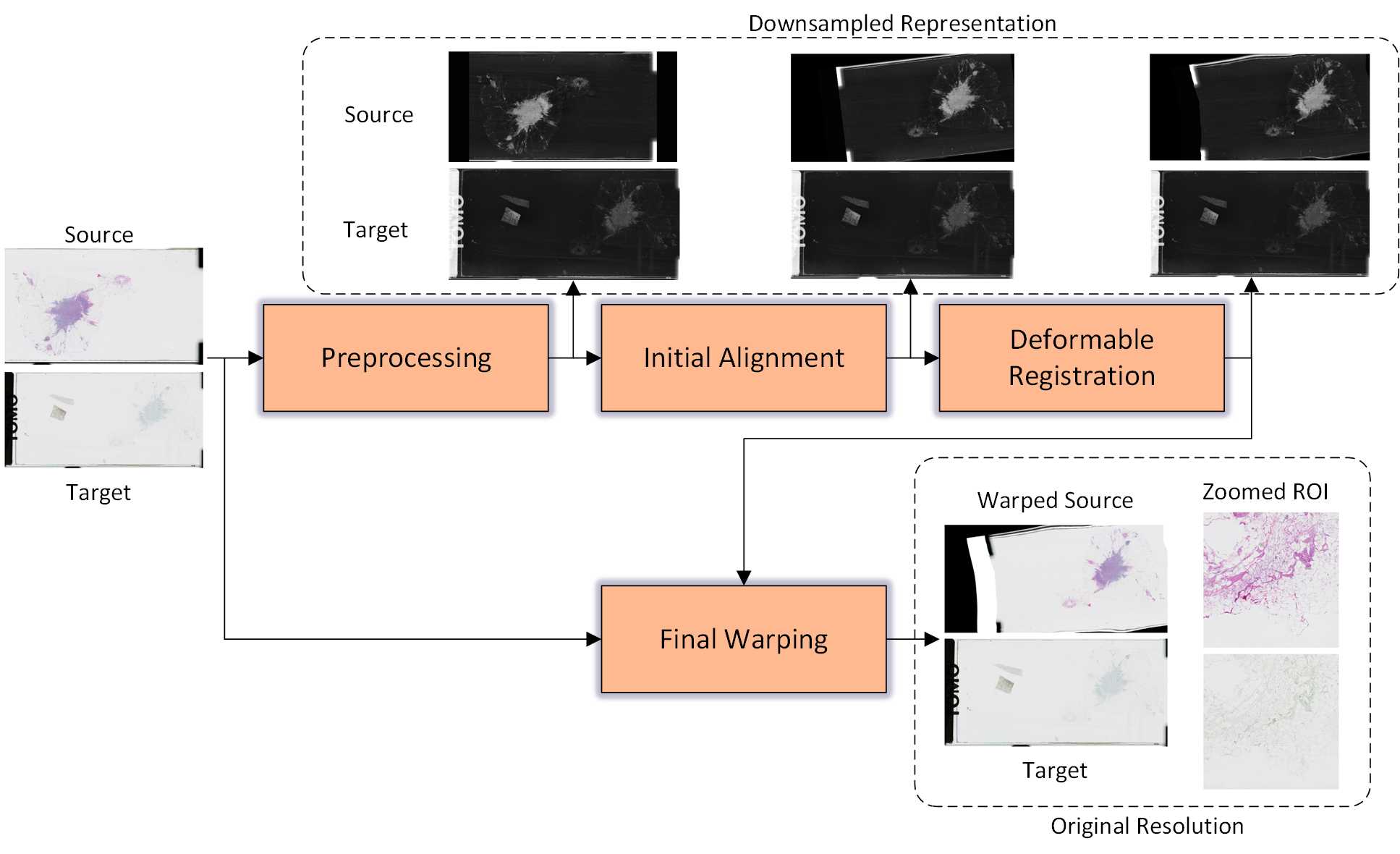}
    \caption{The pipeline of the proposed method. The processing starts with the preprocessing consisting of resampling and color normalization, followed by the robust initial alignment, and finally the accurate deformable registration. The final warping is performed in a separate module at the original resolution.}
    \label{fig:pipeline}
\end{figure*}

\section{Methods}
\label{sec:methods}

\subsection{Overview}

The method consists of three steps: (i) the preprocessing, (ii) the initial alignment, and (iii) the deformable registration. The goal of preprocessing is to convert the input images to the same color space and to resample them to resolutions desired at subsequent registration steps. The initial alignment finds the global affine transformation to roughly match the images. The deformable registration calculates the dense displacement field to recover more complex and nonrigid deformations. All these steps are described in the following sections and presented in Figure~\ref{fig:pipeline}.

\subsection{Preprocessing}

\textcolor{\reviewcolor}{
The preprocessing starts with an automatic estimation of the pyramid level that is then loaded and resampled based on the resolution desired at the nonrigid registration step. The resolution level with the closest higher resolution to the desired one is chosen and then the image is loaded and resampled. Such an approach limits the computational complexity since the initial downsampling is performed using a relatively similar resolution. The initial loading and resampling are implemented using the PyVips~\cite{PyVips} patch-based approach. The initial downsampling is necessary because the input images may be stored in a format not supporting pyramidal representation. For such cases performing the patch-based downsampling to each pipeline step would significantly increase the registration time.}

Next, the images are converted to grayscale, normalized to the [0-1] range, and both the \textcolor{\reviewcolor}{images} are processed with the CLAHE algorithm. Finally, the images are resampled again to the resolutions defined in the configuration of the initial alignment step, preceded by Gaussian smoothing to avoid aliasing. \textcolor{\reviewcolor}{Since the interpolation performs only downsampling and is applied at most two times, the interpolation error is negligible.} 

\subsection{Initial Alignment}

\begin{figure*}[!htb]
    \centering
    \includegraphics[width = 0.95\textwidth]{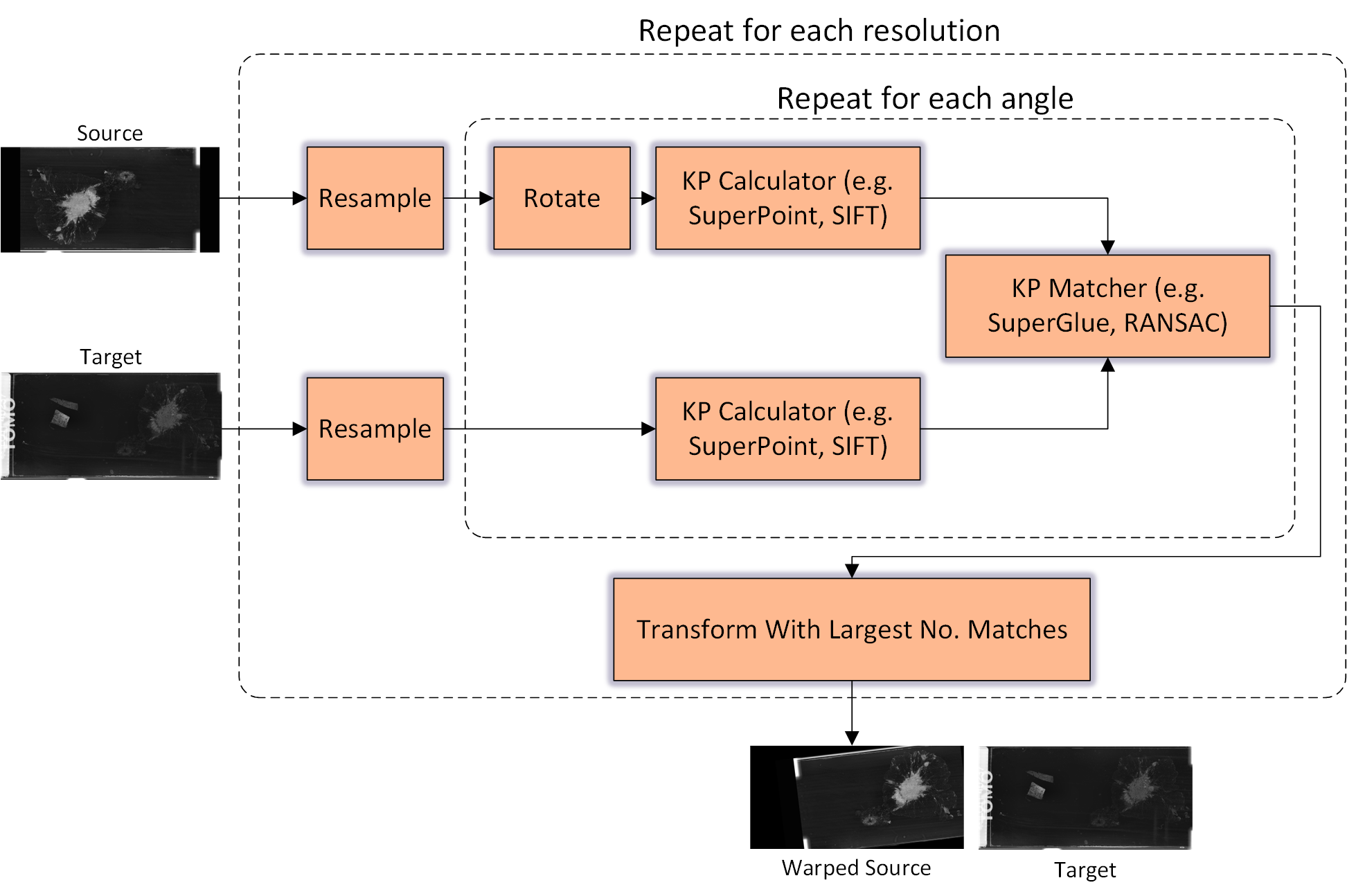}
    \caption{The multi-scale and multi-angle initial alignment pipeline. Since the keypoint extractors are never perfectly scale- and rotation-invariant, such an exhaustive procedure improves the alignment robustness. The transform with the largest number of matches is chosen as the final one, choosing from all transformations calculated across the predefined scales and rotation angles. The procedure improves the alignment quality especially for outliers which would be otherwise incorrectly registered, potentially even decreasing the results quality with respect to the initial misalignment.}
    \label{fig:initial_alignment}
\end{figure*}

The initial alignment is based on pretrained SuperPoint and SuperGlue methods~\cite{detone2018superpoint,sarlin2020superglue}. The SuperPoint is responsible for the feature extraction and the SuperGlue for the feature pairing. The paired features are then used to calculate the affine transformation matrix using the least squares method. Importantly, the SuperPoint and SuperGlue algorithms are not fine-tuned to the histology data. \textcolor{\reviewcolor}{The trained weights based on self-supervised learning are directly} used to calculate and match the features. It confirms the superior generalizability of the SuperPoint feature extractor.

Nevertheless, the SuperPoint and SuperGlue methods are not orientation and scale invariant. Therefore, we extend these methods by embedding them in a pipeline that firstly resamples the source and target images to a given resolution (to extract features at different scales) and secondly rotates the source image with given angles to improve the robustness with respect to the initial orientation. The feature extraction and pairing are then performed repeatedly for each scale and orientation. The result with the highest number of matches calculated by the SuperGlue is chosen as the final outcome. The exhaustive search for the best scale and initial orientation is crucial to increase the robustness of the initial alignment and to make it useful for various datasets. Even though for high-quality datasets a single scale and orientation may be enough, for cases with poor quality and artifacts the exhaustive approach positively impacts the robustness. The initial alignment procedure is shown in Figure~\ref{fig:initial_alignment}.

\subsection{Nonrigid Registration}

The nonrigid registration is implemented as a multilevel instance optimization-based iterative algorithm. The objective function is defined as the weighted sum of the \textcolor{\reviewcolor}{local} normalized cross-correlation (NCC) as the similarity measure, and the diffusive regularization as the regularizer, defined as:
\begin{equation}
    O_{REG}(S, T, u) = NCC(S_i \circ u_i, T_i) + \theta_i Reg(u_i),
\end{equation}
where $S_i, T_i$ are the source and target slides at the i-th resolution level respectively, $u_i$ is the calculated displacement field, $\theta_i$ denotes the regularization coefficient at i-th resolution level, $NCC$ denotes the normalized cross-correlation, $Reg$ is the diffusive regularization, $\circ$ denotes the warping operation, and $N$ is the number of resolution levels.

\textcolor{\reviewcolor}{
We decided to use the local $NCC$ as the similarity measure because the local contrast changes can be accurately captured by the cross-correlation, unlike as in e.g. CT/MRI registration. Moreover, our recent studies confirmed the great potential of NCC for histology registration, outperforming multimodal metrics like Modality Independent Neighbourhood Descriptor, Mutual Information, or Normalized Gradient Fields~\cite{anhir_evaluation, ACROBAT1, wodzinski2021multistep, wodzinski2021deephistreg}.}

The nonrigid registration starts with warping the source image using the transformation found in the initial alignment step. \textcolor{\reviewcolor}{The warping is applied by transforming the affine matrix into displacement field and performing bicubic interpolation. We have chosen this approach because initializing the displacement field with the affine transformation impacts the diffusive regularization term and expects larger receptive field from the optimizer}. In theory, we could avoid that by using the curvature regularization, however, the initial ablations confirmed that the curvature regularization decreases the results quality and is more prone to folding than the diffusive one. Moreover, the proposed initial alignment step is robust and effective, therefore, it can be enforced as the initial status of the deformable registration. The result of the nonrigid registration is saved and can be used to transform the source image at any desired resolution level by using the B-Splines-based upsampling.

All the hyperparameters are adjustable to each resolution level separately. The relative weight of the similarity measure and regularizer can be set differently for each resolution which makes sense since at higher resolutions the weight of regularization should be increased to avoid tissue folding. The same applies to the learning rate that is higher at coarse resolutions and then decreases when the resolution increases. The number of iterations is predefined in the registration settings, without any early stopping mechanism. \textcolor{\reviewcolor}{We decided to use fixed number of iterations because tuning this parameter would require combined grid search together with an optimal regularization coefficient and learning rate which is beyond the computational power of our infrastructure and evaluation possibilities of the Grand-Challenge platform. Moreover, such hyperparameter tuning would require to be performed separately for each dataset. Therefore, we decided that it is better idea to set the parameter fixed and show the robustness of the proposed method because its accurarcy outperforms other methods without any dedicated hyperparameter adjustment.}

\subsection{Datasets}

The evaluation is performed using three open datasets: (i) Automatic Nonrigid Histological Image Registration Dataset (ANHIR), (ii) Automatic Registration of Breast Cancer Tissue Dataset (ACROBAT), and (iii) Hybrid Restained and Consecutive Histological Serial Sections Dataset (HyReCo). Exemplary registration pairs from each dataset are shown in Figure~\ref{fig:dataset}.

The ANHIR dataset consists of 481 image pairs and is split into 230 training and 251 evaluation pairs \cite{ANHIR1,Dataset1,Dataset2,Dataset3,Dataset4}. Noteworthy, the ANHIR dataset and the evaluation platform are biased and almost half of the dataset has to be excluded from the evaluation. Even though the results in the system are aggregated for both training and evaluation subsets, the annotations for the training set are openly released. Therefore, we exclude all the training pairs \textcolor{\reviewcolor}{(N=230)} from the ANHIR dataset in our evaluation to make a fair comparison. The evaluation is performed server-side, automatically using scripts provided by the ANHIR organizers~\cite{anhir_evaluation}. \textcolor{\reviewcolor}{The inter-observer variability is 0.05\% of the image size which can be used as the indication of the best performance possible for the automatic registration methods, below which the methods become indistinguishable~\cite{ANHIR1}}. The images are provided in .jpg and .png format, without the resolution pyramid and metadata. There are 8 different tissue types which were stained using 10 different dyes. A full description of the dataset can be found in~\cite{ANHIR1}.

The ACROBAT dataset consists of consecutive female breast cancer slides. There are 750, 100, and 303 training, validation, and test cases respectively. Since the cases are stained using Hematoxylin\&Eosin (H\&E), ER, KI67, PGR, and HER2, there are 3406 training, validation, and test WSIs respectively (the training cases are stained using multiple dyes). The WSIs were acquired using three scanners, one NanoZoomer S360, and two different NanoZoomer XRs. The images are saved and released in .tiff format using 10X magnification, downscaled from the original 40X. The dataset was annotated by 13 experienced individuals on the original 40X scale. In this paper, the evaluation is performed on the 100 hidden validation pairs using the ACROBAT evaluation platform~\cite{acrobat_evaluation}. Unfortunately, the 303 test cases are hidden and not released publicly. The description of the ACROBAT dataset is available in~\cite{ACROBAT1,ACROBAT2}.

The HyReCo dataset, released in 2021, is unique by providing both consecutive and restained sections. The dataset was acquired at the Radboud University Medical Center and annotated by experienced researchers from Fraunhofer MEVIS. The dataset consists of 9 high-resolution consecutive sections, each stained with H\&E, CD8, CD45, and KI67. For each section, there are between 11 and 19 annotated landmarks, resulting in 138 annotations per stain and 690 annotations in total. The landmarks were placed and verified by two experienced researchers. \textcolor{\reviewcolor}{Additional slides} were restained using the PHH3, resulting in 54 restained H\&E/PHH3 pairs with about 43 annotations per restained pair. The slides are saved and released in BigTIFF format with corresponding landmarks released in a CSV file containing world coordinates in mm. The HyReCo dataset, together with its full description, is available in~\cite{HyReCo1,HyReCo2}.

\begin{figure*}[!htb]
    \centering
    \includegraphics[width = 0.95\textwidth]{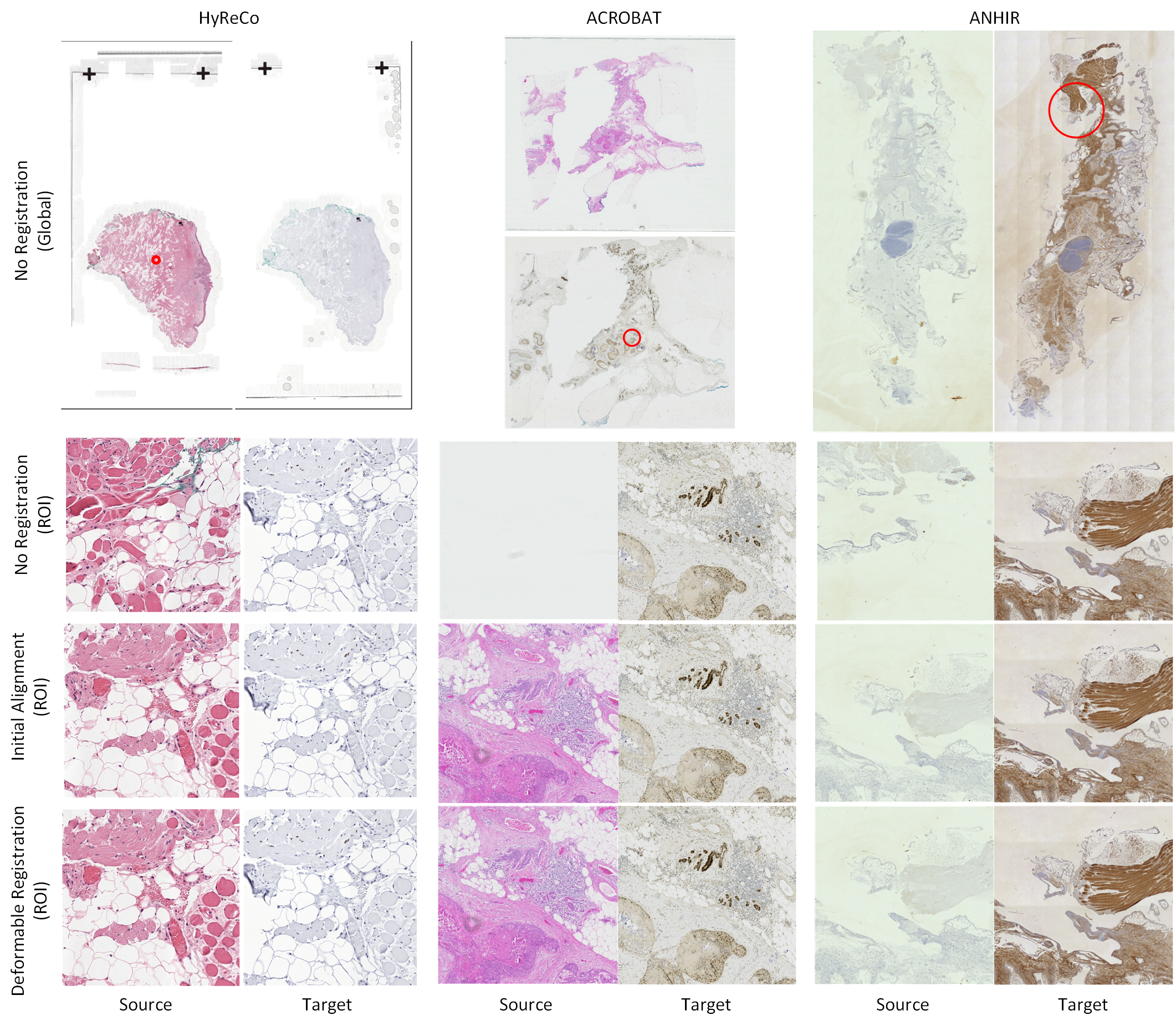}
    \caption{Visual registration results for exemplary cases from all three datasets. The results are presented on a chosen ROI after subsequent registration steps. The global overview for ANHIR case is rotated, and the ACROBAT source image is cropped, for the presentation clarity. Note the difference between the registration quality of restained (HyReCo) and consecutive slides (ACROBAT/ANHIR).}
    \label{fig:visual_results}
\end{figure*} 

\begin{figure*}[!htb]
    \centering
    \includegraphics[width = 0.95\textwidth]{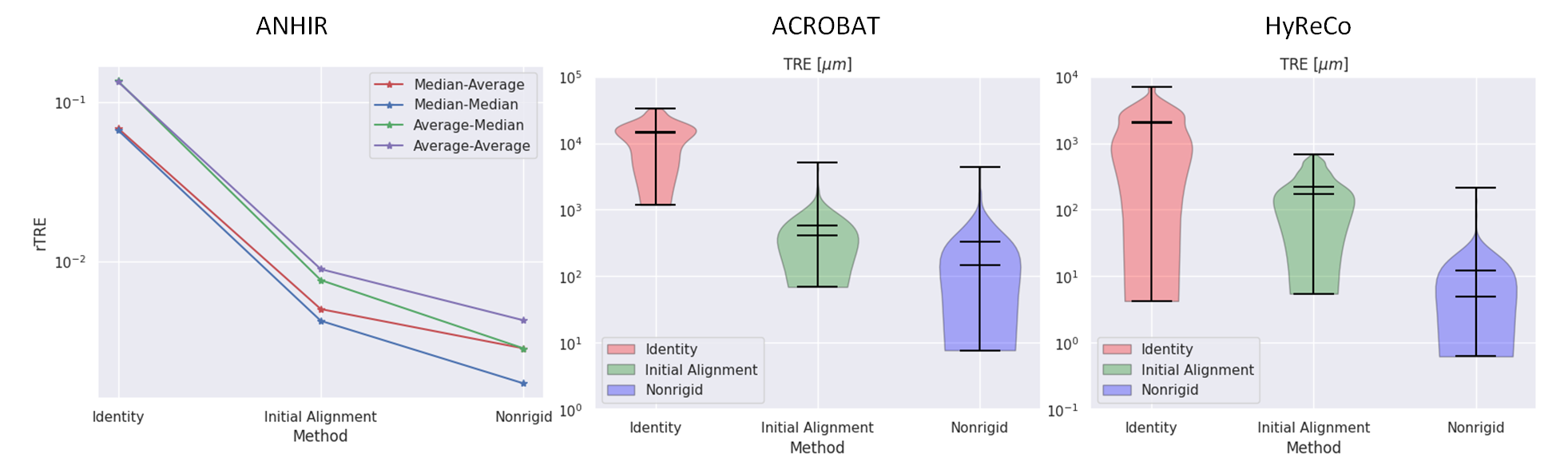}
    \caption{The quantitative results in terms of the target registration error (TRE) after the subsequent registration steps. The HyReCo results are reported for the consecutive slides only. \textcolor{\reviewcolor}{Note the significant improvement (p-value < 0.01) after each registration step.}}
    \label{fig:steps}
\end{figure*}

\subsection{Experimental Setup}
We compare the proposed method to the state-of-the-art solutions. The target registration error (TRE) is used as the evaluation metric. The TRE calculates the Euclidean distance between corresponding landmarks, manually annotated by experts. For the ACROBAT and HyReCo datasets, we report the TRE in physical units while for the ANHIR dataset, we normalize it by diving by the image diagonal (rTRE), following the convention introduced by the \textcolor{\reviewcolor}{ANHIR dataset creators and other methods reporting the results~\cite{ANHIR1,ANHIR2}.}

We perform several ablation studies to confirm the method's robustness and the impact of the most crucial hyperparameters. The experiments verify (i) the influence of each registration step, (ii) the difference between the SuperPoint/SuperGlue method and the traditional SIFT/RANSAC approach~\cite{wodzinski2021multistep}, (iii) the influence of the number of rotation angles and resolutions on the initial alignment stability, (iv) the impact of the resolution on the nonrigid registration quality, both for the consecutive and restained slides. \textcolor{\reviewcolor}{Any claim in the Results or Discussion sections about significant improvement means that the claim is supported by Wilcoxon signed-rank test with p-value lower than 0.01.}

All the algorithms and experiments are implemented using the PyTorch library with extensions based on PyVips~\cite{PyVips}. The experiments were performed using NVIDIA RTX 3090 GPU with 24 GB of dedicated memory. The proposed method is integrated into the DeeperHistReg framework and openly released~\cite{deeperhistreg}. The method does not require any fine-tuning or retraining to be used for other datasets.

\section{Results}
\label{sec:results}

\subsection{Affine vs Nonrigid Registration}

Figure~\ref{fig:steps} presents the mean and median TRE distribution after each registration step. The ticks in violin plots present the mean and median value respectively. It can be noted that the the registration improves the alignment for majority of registration pairs, without a single case that could be considered a failure (for which mean/median TRE significantly increases after the registration). \textcolor{\reviewcolor}{The Median-Average rTRE before the registration, after the affine alignment, and after the nonrigid registration are 0.0683, 0.0050, 0.0029 for the ANHIR dataset. For the ACROBAT dataset, the Median TRE in $\mu$m are 14424.35, 409.86, 137.27 after subsequent registration steps. Finally, for the HyReCo datasets, the Median TRE in $\mu$m are 174.90, 4.96 (consecutive), and 12.24, 0.59 (restained) after the initial alignment and the deformable registration, respectively. It confirms significant improvement after each registration step (p-value < 0.01).}  Exemplary visualization of the registration after each algorithm step is shown in Figure~\ref{fig:visual_results}. Importantly, the percentage of foldings, defined as the ratio of a number of pixels with negative Jacobian to all pixels is below 0.1\% for all nonrigid experiments.

\subsection{Comparison between SuperPoint-SuperGlue and SIFT-RANSAC}

\begin{figure}[!htb]
    \centering
    \includegraphics[width = 0.45\textwidth]{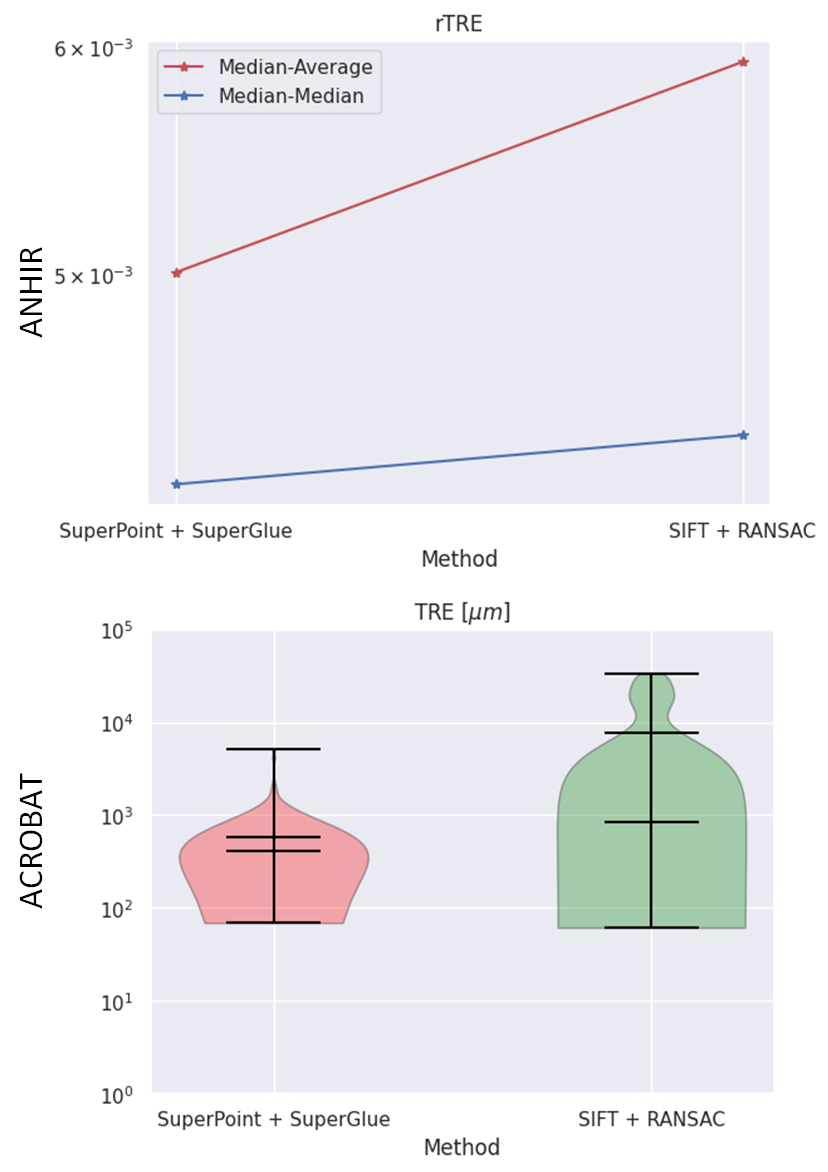}
    \caption{Comparison of the SuperPoint/SuperGlue-based approach to the SIFT/RANSAC method. The Average-Median and Average-Average for ANHIR dataset are skipped due to large number of failures of the SIFT/RANSAC method (more than 10\%). The comparison is skipped for the HyReCo dataset due to large number of failures of SIFT/RANSAC method (more than 10\%).}
    \label{fig:sp_sr}
\end{figure}

One of the ablations confirms the impact of the novel SuperPoint/SuperGlue initial registration to the state-of-the-art SIFT/RANSAC methods. The results are presented in Figure~\ref{fig:sp_sr}. One can note that the learning-based methods are more robust without failures which are quite common for the SIFT/RANSAC method. \textcolor{\reviewcolor}{The Median-Average rTRE for SuperPoint/SuperGlue, and SIFT + RANSAC are equal to 0.0050, 0.0059 for the ANHIR dataset. For the ACROBAT dataset, the Median-Average TRE in $\mu$m for those methods are equal to 409.86, 852.43. The comparison is skipped for the HyReCo dataset due to large number of failrues of SIFT/RANSAC method accounting for more than 10\% of the registered pairs.} Moreover, the proposed learning-based method is computationally more efficient ($\sim$15 vs $\sim$27 seconds), however, the effect is negligible when considering the time required for further nonrigid registration and final image warping.

\subsection{Stability concerning the number of angles and scales}

\textcolor{\reviewcolor}{The outcomes of the experiment related to the influence of the number of scales are shown in Table~\ref{tab:ia_scales}, and the impact of the rotation angle step is shown in Table~\ref{tab:ia_rotation_angles}. The difference between using a single scale and using eight scales is notable (0.0128 vs 0.0057 in terms of the rTRE for the ANHIR dataset, 470.98 vs 396.38 $\mu$m for the ACROBAT dataset, and 288.47 vs 232.64 $\mu$m for the HyReCo dataset). The small angle step (30 degrees vs 180 degrees) is also influential (0.0057 vs 0.0110 in rTRE for the ANHIR dataset, 396.38 vs 417.30 $\mu$m for the ACROBAT dataset), with the exception of HyReCo datasets where the best alignment was achieved for angle step equal to 120. The differences in terms of both Average-Median and Average-Average TRE are significant for both angle- and scale-related ablations (p-value < 0.01) for all the datasets. It can be noted that the multi-scale and multi-angle extension is not crucial to achieving considerable performance, better than the multi-scale/multi-angle SIFT/RANSAC method. However, it ensures as high robustness as possible, without failures of the initial alignment.}

\begin{table}[!htb]
\centering
\caption{The effect of the number of scales on the stability of the initial alignment based on the SuperPoint and SuperGlue descriptors. In case of the HyReCo dataset, the results are reported for the consecutive slides.}
\renewcommand{\arraystretch}{1.0}
\footnotesize
\resizebox{0.5\textwidth}{!}{%
\begin{tabular}{ccc}
\label{tab:ia_scales}
\tabularnewline
\multicolumn{3}{c}{ANHIR}
\tabularnewline
\multicolumn{1}{c}{$\#$Scales} & \multicolumn{1}{c}{Average-Median rTRE} & \multicolumn{1}{c}{Average-Average rTRE}
\tabularnewline
\hline
1 & 0.0128 & 0.0140 \tabularnewline
2 & 0.0092 & 0.0104 \tabularnewline
3 & 0.0083 & 0.0095 \tabularnewline
4 & 0.0068 & 0.0080 \tabularnewline
5 & 0.0063 & 0.0074 \tabularnewline
6 & 0.0061 & 0.0072 \tabularnewline
7 & 0.0058 & \textbf{0.0070} \tabularnewline
8 & \textbf{0.0057} & 0.0071 \tabularnewline
\multicolumn{3}{c}{ACROBAT}
\tabularnewline
\multicolumn{1}{c}{$\#$Scales} & \multicolumn{1}{c}{Average-Median TRE [$\mu$m]} & \multicolumn{1}{c}{Average-Average TRE [$\mu$m]}
\tabularnewline
\hline
1 & 470.98 & 656.73 \tabularnewline
2 & 452.78 & 946.40 \tabularnewline
3 & 434.55 & 629.83 \tabularnewline
4 & 466.00 & 614.22 \tabularnewline
5 & 446.39 & 562.28 \tabularnewline
6 & 410.68 & 541.34 \tabularnewline
7 & 404.90 & \textbf{537.69} \tabularnewline
8 & \textbf{396.38} & 540.87 \tabularnewline
\multicolumn{3}{c}{HyReCo}
\tabularnewline
\multicolumn{1}{c}{$\#$Scales} & \multicolumn{1}{c}{Average-Median TRE [$\mu$m]} & \multicolumn{1}{c}{Average-Average TRE [$\mu$m]}
\tabularnewline
\hline
1 & 288.47 & 321.82 \tabularnewline
2 & 275.28 & 312.13 \tabularnewline
3 & 289.18 & 323.75 \tabularnewline
4 & 255.69 & 291.04 \tabularnewline
5 & 263.44 & 298.31 \tabularnewline
6 & 238.13 & 268.99 \tabularnewline
7 & 234.92 & 266.39 \tabularnewline
8 & \textbf{234.64} & \textbf{263.08} \tabularnewline
\hline
\end{tabular}}

\end{table}

\begin{table}[!htb]
\centering
\caption{The effect of the rotation angle step on the stability of the initial alignment based on the SuperPoint and SuperGlue descriptors. In case of the HyReCo dataset, the results are reported for the consecutive slides. Note that the results for the HyReCo dataset are strongly influenced by the fact that in several slides there are more than one tissue.}
\renewcommand{\arraystretch}{1.0}
\footnotesize
\resizebox{0.5\textwidth}{!}{%
\begin{tabular}{ccc}
\label{tab:ia_rotation_angles}
\tabularnewline
\multicolumn{3}{c}{ANHIR}
\tabularnewline
\multicolumn{1}{c}{Angle Step} & \multicolumn{1}{c}{Average-Median rTRE} & \multicolumn{1}{c}{Average-Average rTRE}
\tabularnewline
\hline
30 & \textbf{0.0057} & \textbf{0.0071} \tabularnewline
60 & 0.0079 & 0.0093 \tabularnewline
90 & 0.0079 & 0.0093 \tabularnewline
120 & > 0.1 & > 0.1 \tabularnewline
150 & 0.0084 & 0.0098 \tabularnewline
180 & 0.0110 & 0.0124 \tabularnewline

\multicolumn{3}{c}{ACROBAT}
\tabularnewline
\multicolumn{1}{c}{Angle Step} & \multicolumn{1}{c}{Average-Median TRE [$\mu$m]} & \multicolumn{1}{c}{Average-Average TRE [$\mu$m]}
\tabularnewline
\hline
30 & \textbf{396.38} & \textbf{540.87} \tabularnewline
60 & 409.86 & 574.50 \tabularnewline
90 & 401.58 & 546.31 \tabularnewline
120 & 490.42 & 993.30 \tabularnewline
150 & 401.587 & 588.47 \tabularnewline
180 & 417.30 & 561.56 \tabularnewline

\multicolumn{3}{c}{HyReCo}
\tabularnewline
\multicolumn{1}{c}{Angle Step} & \multicolumn{1}{c}{Average-Median TRE [$\mu$m]} & \multicolumn{1}{c}{Average-Average TRE [$\mu$m]}
\tabularnewline
\hline
30 & 234.64 & 263.08 \tabularnewline
60 & 223.58 & 253.28 \tabularnewline
90 & 223.58 & 253.28 \tabularnewline
120 & \textbf{52.45} & \textbf{65.18} \tabularnewline
150 & 138.85 & 156.22 \tabularnewline
180 & 223.58 & 253.28 \tabularnewline

\hline
\end{tabular}}

\end{table}

\subsection{Impact of the resolution on consecutive and restained slices}

\begin{figure*}[!htb]
    \centering
    \includegraphics[width = 0.85\textwidth]{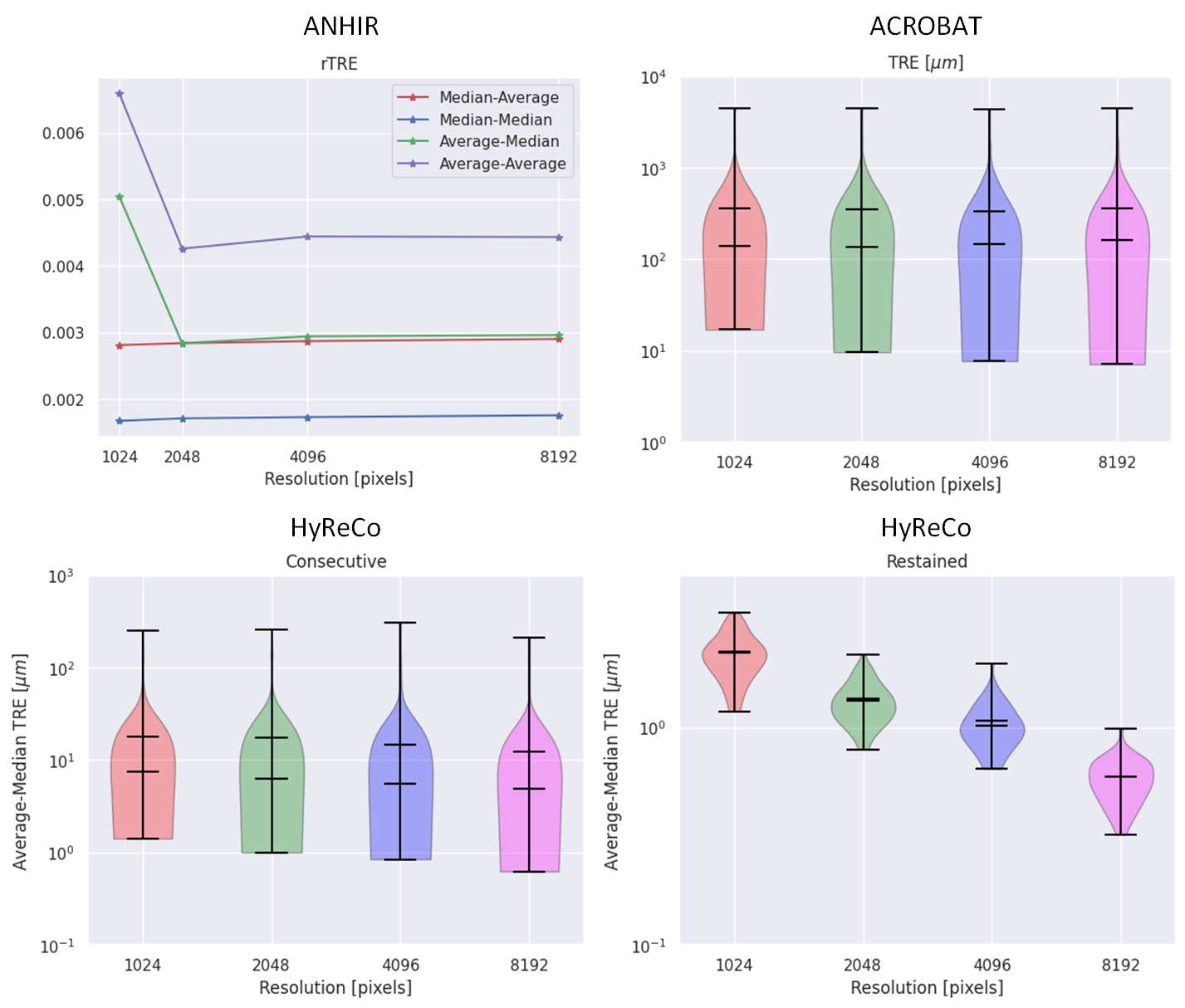}
    \caption{The influence of the image resolution on the deformable registration quality. Note that for the consecutive slides increasing the resolution has negligible positive effect (ACROBAT, HyReCo consecutive) or even decreases the performance (ANHIR), however, \textcolor{\reviewcolor}{for the restained ones increasing the resolution has a significant impact (p-value < 0.01) on the registration quality.}}
    \label{fig:resolution}
\end{figure*}

\textcolor{\reviewcolor}{The influence of the registration resolution on the nonrigid registration step is shown in Figure~\ref{fig:resolution} for each dataset. The Figure~\ref{fig:resolution} also presents the comparison between the consecutive and the restained slices from the HyReCo dataset. The comparison between the consecutive and restained slides is reported only for the HyReCo cases because other datasets consist of only consecutive slides. It can be noted that the resolution used for registration is less impactful for the consecutive slides and easily saturates (insignificant changes), however, for the restained ones further increase of the resolution continues to significantly (p-value < 0.01) improve the registration quality (MTRE equal to 2.20, 1.33, 1.01, 0.59 $\mu$m for $1024^2$, $2048^2$, $4096^2$, $8192^2$ respectively). Therefore, the increase of the registration resolution beyond $8192^2$ would be beneficial only for the restained slides, however, such a change would require a patch-based approach to the registration due to the GPU memory limitations.}

\subsection{Comparison to the state-of-the-art}

\begin{figure}[!htb]
    \centering
    \includegraphics[width = 0.48\textwidth]{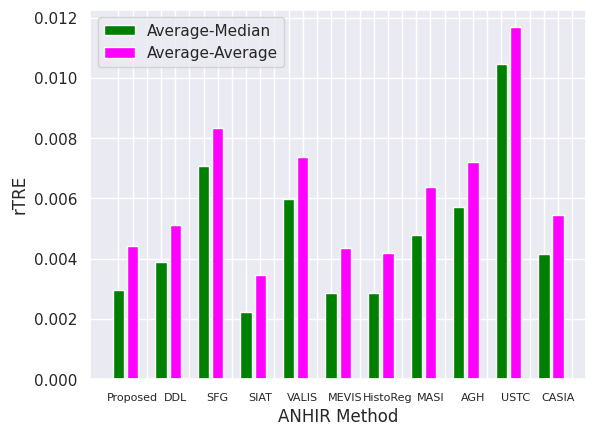}
    \caption{Comparison to the best performing methods on the evaluation subset of the ANHIR dataset. The results are taken directly from the ANHIR evaluation platform for the best submission from each participant. The challenge platform reports only the aggregates.}
    \label{fig:anhir_comparison}
\end{figure}

\begin{figure}[!htb]
    \centering
    \includegraphics[width = 0.48\textwidth]{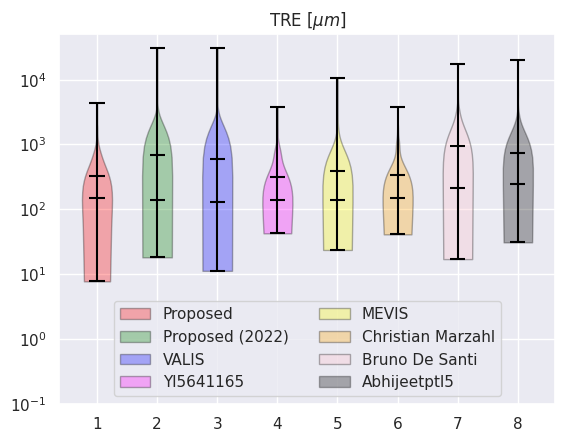}
    \caption{Comparison to the best performing methods on the hidden validation subset of the ACROBAT dataset. The results are taken directly from the ACROBAT evaluation platform for the best submission from each participant.}
    \label{fig:acrobat_comparison}
\end{figure}

\textcolor{\reviewcolor}{
The comparison to the state-of-the-art methods using the ANHIR and ACROBAT datasets are shown in Table~\ref{tab:comparison_anhir} and Figures~\ref{fig:anhir_comparison},~\ref{fig:acrobat_comparison} respectively. The results for the state-of-the-art methods are acquired by downloading the best submission for each method from the ANHIR/ACROBAT challenge platforms. For the ANHIR dataset, we report results only using the evaluation cases because the annotations for the training set are openly available which biases towards the learning-based methods utilizing the annotations during training. The proposed method achieves only slightly worse results compared to the best-performing ones (0.0017 vs 0.0015 for Median-Median rTRE and 0.0044 vs 0.0034 for the Average-Average rTRE). For the ACROBAT dataset, we report the results using the closed validation set. The method has the lowest Average-Median TRE among all the proposed methods (331.01 $\mu$m vs 385.58 $\mu$m achieved by HistoKat Fusion, the second most accurate registration method.). For the HyReCo dataset, we compare the proposed method to the HistokatFusion in Table~\ref{tab:comparison_hyreco} by using the results reported in the respective publication~\cite{HyReCo1}. The method outperforms the HistokatFusion in the deformable registration (0.90$\mu$m, 4.96$\mu$m for the restained, and consecutive slides respectively, versus 0.90 $\mu$m and 5.30 $\mu$m achieved by the HistoKat Fusion).}

\begin{table}[!htb]
\centering
\caption{Comparison of the proposed method to other best-performing methods evaluated on the ANHIR dataset. The results are taken directly from the ANHIR evaluation platform for the best submission from each participant. The results are reported for the evaluation subset of the ANHIR dataset. The Med-Med and Med-Avg show the ability of the methods to recover fine details while the Avg-Med and Avg-Avg present the robustness with respect to difficult cases.}
\renewcommand{\arraystretch}{1.0}
\footnotesize
\resizebox{0.5\textwidth}{!}{%
\begin{tabular}{ccccc}
\label{tab:comparison_anhir}
Method & \multicolumn{1}{c}{Med-Med rTRE $\downarrow$} & \multicolumn{1}{c}{Med-Avg rTRE $\downarrow$}  &
\multicolumn{1}{c}{Avg-Med rTRE $\downarrow$} &
\multicolumn{1}{c}{Avg-Avg rTRE $\downarrow$} 
\tabularnewline
\hline
Proposed & 0.0017 & 0.0029 & 0.0029 & 0.0044
\tabularnewline
DDL & \textbf{0.0015} & \textbf{0.0023} & 0.0039 & 0.0051
\tabularnewline
SFG & 0.0016 & 0.0025 & 0.0070 & 0.0083
\tabularnewline
SIAT & 0.0016 & 0.0024 & \textbf{0.0022} & \textbf{0.0034}
\tabularnewline
VALIS & 0.0017 & 0.0025 & 0.0059 & 0.0073
\tabularnewline
MEVIS & 0.0018 & 0.0027 & 0.0029 & 0.0043
\tabularnewline
HistoReg & 0.0019 & 0.0029 & 0.0029 & 0.0042
\tabularnewline
MASI & 0.0021 & 0.0031 & 0.0048 & 0.0064
\tabularnewline
AGH & 0.0017 & 0.0031 & 0.0057 & 0.0071
\tabularnewline
USTC & 0.0024 & 0.0037 & 0.0104 & 0.0117
\tabularnewline
CASIA & 0.0026 & 0.0037 & 0.0041 & 0.0054
\tabularnewline

\hline

\hline
\end{tabular}}

\end{table}
\begin{table}[!htb]
\centering
\caption{Comparison of the proposed method to the HistokatFusion on the HyReCo consecutive and restained slides in terms of the median TRE (MTRE). Note that even though the initial alignment is worse than for the HistokatFusion, the following deformable refinement still significantly improves the results.}
\renewcommand{\arraystretch}{1.0}
\footnotesize
\resizebox{0.5\textwidth}{!}{%
\begin{tabular}{ccccc}
\label{tab:comparison_hyreco}
Method & \multicolumn{2}{c}{MTRE Affine [$\mu$m] $\downarrow$} & \multicolumn{2}{c}{MTRE Deformable [$\mu$m] $\downarrow$} \tabularnewline
\multicolumn{1}{c}{} & \multicolumn{1}{c}{Restained} & \multicolumn{1}{c}{Consecutive} & \multicolumn{1}{c}{Restained} & \multicolumn{1}{c}{Consecutive}
\tabularnewline
\hline
\multicolumn{1}{c}{Proposed} & \multicolumn{1}{c}{12.24} & \multicolumn{1}{c}{174.90} & \multicolumn{1}{c}{\textbf{0.59}} & \multicolumn{1}{c}{\textbf{4.96}}
\tabularnewline
\multicolumn{1}{c}{HistokatFusion~\cite{HyReCo1}} & \multicolumn{1}{c}{\textbf{1.70}} & \multicolumn{1}{c}{\textbf{20.20}} & \multicolumn{1}{c}{0.90} & \multicolumn{1}{c}{5.30} \tabularnewline
\hline
\end{tabular}}
\end{table}

\section{Discussion}
\label{sec:discussion}

The proposed method produces accurate registrations for all three datasets, as shown in Figure~\ref{fig:steps}. Importantly, the method uses the same configuration for all registration pairs and does not require adjusting the parameters to a particular dataset. The method is the most accurate on the ACROBAT benchmark, among the best-performing ones for the ANHIR dataset, and comparable to the HistokatFusion on the HyReCo slides. The average registration time, for the $8192^2$ resolution is below 2 minutes.

Interesting observations, confirming the results obtained by the creators of the HyReCo dataset~\cite{HyReCo1}, can be made when comparing the registration accuracy between the consecutive and restained slides (Figure~\ref{fig:resolution}). Increasing the registration resolution substantially decreases the TRE for the restained slides, however, for the consecutive ones the impact is less significant, and further increase of the resolution does not improve the results. Even though the nonrigid registration still significantly improves the accuracy when compared to the initial affine alignment, one should keep in mind that achieving the cell-level registration accuracy for the consecutive slides may be challenging if not impossible. The registration accuracy strongly depends on the quality of the slides, therefore to achieve as accurate registration as possible, it is important to prepare the slides carefully.

The initial alignment method is stable and robust, as presented in Figure~\ref{fig:steps} and Tables~\ref{tab:ia_scales},~\ref{tab:ia_rotation_angles}. The failures are not significant for the HyReCo dataset, and there are no failures for the ANHIR and ACROBAT datasets. The initial alignment decreases the initial TRE for all registration pairs in ANHIR and ACROBAT datasets and increases the TRE only those pairs in the HyReCo dataset that contain multiple tissue sections. Tabels~\ref{tab:ia_scales},~\ref{tab:ia_rotation_angles} confirm the impact of the multi-scale and multi-angle approach. Even though it is not necessary for high-quality slides, it improves the stability of the low-quality ones. The results also show that the tissues are usually rotated by 0, 45, 90, 135, or 180 degrees. The experiments with the angle step equal to 60 or 120 degrees resulted in several outliers. One could argue that the performance of the initial alignment for the HyReCo dataset is suboptimal, however, it turns out that it does not impact the following deformable step for both consecutive and restained slides. Moreover, the suboptimal performance is caused by factors that should be addressed during the quality control step.

The algorithm has several advantages. The background segmentation~\cite{bandi2019resolution,jurgas2023robust}, which is a preprocessing step for several other contributions, is not necessary. The need for background segmentation usually decreases the registration methods' robustness and generalizability because the segmentation is nowadays based on learning-based techniques. As far as Hematoxylin\&Eosin (H\&E) images are concerned, the problem may be considered solved, however, the current state-of-the-art methods still struggle with other stains and need further improvement. Therefore, eliminating the necessity of background segmentation by using the feature-based approach to the initial alignment increases the robustness and eliminates the error-prone part of the intensity-based pipeline. 

Another advantage is connected with the method's high robustness and generalizability. Even though parts of the method are learning-based, they do not require fine-tuning to new datasets. The instance optimization approach by design captures the heterogeneity of WSIs by using the inference time optimization and the SuperPoint/SuperGlue methods are characterized by high stability and reliability. 

We openly release the method's source code and incorporate it in the DeeperHistReg framework allowing one to reproduce the results and use the method in their research. The method is ready to use, one needs to only clone the repository, set up paths to the images to be registered, and run the registration. This is a significant contribution to the field of digital pathology since other registration methods are often closed source or have a high entry barrier requiring the registration expertise from the end-user. Importantly, the DeeperHistReg framework allows one to not only perform the registration but also to load any OpenSlide compatible format, transform the WSIs, and save the output at any desired resolution level. Our other experiments confirmed also the usability to other microscopic images, e.g. unstained, DAPI-stained, or ExM images, without the need for the parameters adjustment.

The method has also several limitations. The disadvantage of the deformable step is its limited ability to recover large non-rigid displacements, as confirmed in Table~\ref{tab:comparison_anhir}. It is connected with the limitations of instance optimization that inherently has a narrow receptive field, even in multilevel scenarios. As far as good quality samples are concerned, especially restained ones, the effect is negligible, as presented in Table~\ref{tab:comparison_hyreco}, where the proposed method even outperformed the HistokatFusion. However, for low-quality consecutive slides, the power to recover large deformations could further improve the registration quality, thus further increasing the registration quality for ANHIR and ACROBAT datasets. \textcolor{\reviewcolor}{Another limitation of the method is lack of downstream validation on classification or segmentation tasks, however, this is a topic for separate studies. Additionally, the method could benefit from a complete GPU-based implementation, starting from loading the images directly to the GPU memory and then performing all the steps using just one device. Such an approach would not change the registration accuracy, however, would be beneficial from the practical point of view. Nevertheless, the transition to a GPU-only implementation is a pure technical contribution with limited scientific value.}

One could argue that the proposed modifications to the initial alignment based on the SuperPoint and SuperGlue architectures add additional computational complexity. It is indeed true, however, the time required for the initial alignment is negligible when compared to the time required to perform the following nonrigid registration or to warp the source image after the registration. The initial alignment takes no more than 20 seconds while warping the source image after the registration at the highest resolution level requires between 2 to 10 minutes.

In future work, we plan to extend the method in several ways. First, we forecast that fine-tuning the SuperPoint and SuperGlue methods using self-supervised learning on large histology datasets, e.g. from the BigPicture consortium~\cite{bigpicture} could further improve the robustness of the initial alignment, and allow one to skip the multi-scale and multi-orientation steps. Additionally, it would be interesting to propose variants of the SuperPoint and SuperGlue methods for volumetric data which could be beneficial for the initial alignment in radiology~\cite{hering2022learn2reg}. Another direction we plan to explore is the use of stain normalization and transfer to improve the preprocessing step. The normalization of color spaces of both the source and target images could strongly influence the quality of the deformable registration. Finally, the difficulties of the nonrigid registration to handle complex deformations could be solved by the neural instance optimization of deep models dedicated to recovering large displacements, e.g. LapIRN~\cite{mok2020large,mok2023deformable}.

To conclude, we proposed an automatic, robust, and accurate method dedicated to WSI registration. We evaluated the method using three open datasets: ANHIR, ACROBAT, and HyReCo, performed several ablation studies, and compared it to other state-of-the-art methods. We freely released the source code and incorporated it in the DeeperHistReg framework, allowing other researchers to use the method in their research and to easily reproduce the results. The proposed method is a significant contribution to the field of digital pathology.

%%%%%%%%%%%% Supplementary Methods %%%%%%%%%%%%

%%%%%%%%%%%%% Acknowledgements %%%%%%%%%%%%%
\section*{Acknowledgements}
This project has received funding from the Innovative Medicines Initiative 2 Joint Undertaking under grant agreement No 945358. This Joint Undertaking receives support from the European Union's Horizon 2020 research and innovation program and EFPIA, Belgium (www.imi.europe.eu). The research reflects only the author's view and the Joint Undertaking is not responsible for any use that may be made of the information it contains. Additionally, the research was supported in part by PLGrid Infrastructure. We gratefully acknowledge Poland’s high-performance computing infrastructure PLGrid (HPC Centers: ACK Cyfronet AGH) for providing computer facilities and support within computational grant no. PLG/2023/016239.

%%%%%%%%%%%%%%   Bibliography   %%%%%%%%%%%%%%
\normalsize
\bibliography{references}

%%%%%%%%%%%%  Supplementary Figures  %%%%%%%%%%%%
%\clearpage

%%%%%%%%%%%%%%%%   End   %%%%%%%%%%%%%%%%
%\end{multicols}  % Method B for two-column formatting (doesn't play well with line numbers), comment out if using method A
\end{document}